\documentclass[5p,preprint]{elsarticle}
\biboptions{sort&compress}
\usepackage{lineno,isotope,mathrsfs}\modulolinenumbers[5]
\usepackage{amsmath,amssymb,mathptmx,epsfig,bm,mathrsfs,feynmp}
\usepackage{color,slashed,nicefrac,exscale,multirow,soul}
\usepackage{times,txfonts,xcolor,graphicx,gensymb,tikz}
\usepackage[normalem]{ulem}
\usepackage[breaklinks,colorlinks,citecolor=blue]{hyperref}
\definecolor{red}{rgb}{0.8,0,0}
\definecolor{violet}{rgb}{0.4,0,0.4}
\definecolor{green}{rgb}{0,0.5,0.0}
\definecolor{navy}{rgb}{0.0,0.0,0.6}
\definecolor{orange}{rgb}{0.8,0.2,0.0}

\usepackage[normalem]{ulem} 

\newcommand{\bea}{\begin{eqnarray}}
\newcommand{\eea}{\end{eqnarray}}

\newcommand{\Lsym}{\ensuremath{L_{\text{sym}}}}
\newcommand{\Qsat}{\ensuremath{Q_{\text{sat}}}}
\begin{document}
\begin{frontmatter}
\title{Massive relativistic compact stars from SU(3) symmetric quark models}
\author[a]{Han Rui Fu}
\author[a]{Jia Jie Li}
\ead{jiajieli@swu.edu.cn}
\author[b,c]{Armen Sedrakian}
\ead{sedrakian@fias.uni-frankfurt.de}
\author[d,e]{Fridolin Weber}
\ead{fweber@sdsu.edu}
\address[a]{School of Physical Science and Technology, Southwest University, Chongqing 400700, China}
\address[b]{Frankfurt Institute for Advanced Studies, D-60438
Frankfurt am Main, Germany}
\address[c]{Institute of Theoretical Physics, University of Wroclaw,
50-204 Wroclaw, Poland}
\address[d]{Department of Physics, San Diego State University,
5500 Campanile Drive, San Diego, California 92182, USA}
\address[e]{Center for Astrophysics and Space Sciences,
University of California at San Diego, La Jolla, California 92093, USA}
\begin{abstract} 
  We construct a set of hyperonic equations of state (EoS) by
  assuming SU(3) symmetry within the baryon octet and by using a
  covariant density functional (CDF) theory approach. The low-density
  regions of our EoS are constrained by terrestrial experiments, while
  the high-density regime is modeled by systematically varying the
  nuclear matter skewness coefficient $\Qsat$ and the symmetry energy
  slope $\Lsym$. The sensitivity of the EoS predictions is explored in
  terms of $z$ parameter of the SU(3) symmetric model that modifies
  the meson-hyperon coupling constants away from their SU(6) symmetric
  values.  Our results show that model EoS based on our approach can
  support static Tolman-Oppenheimer-Volkof (TOV) masses in the range 
  $2.3$-$2.5\,M_{\odot}$ in the large-$\Qsat$ and small-$z$ regime, 
  however, such stars contain only a trace amount of hyperons compared 
  to SU(6) models.  We also construct uniformly rotating Keplerian 
  configurations for our model EoS for which the masses of stellar 
  sequences may reach up to $3.0\,M_{\odot}$.  These results are used 
  to explore the systematic dependence of the ratio of maximum masses 
  of rotating and static stars, the lower bound on the rotational frequency 
  of the models that will allow secondary masses in the gravitational 
  waves events to be compact stars with $M_2 \lesssim 3.0\,M_{\odot}$ and 
  the strangeness fraction on the model parameters. We conclude that
   very massive stellar models can be, in principle, constructed
  within the SU(3) symmetric model, however, they are 
  nucleonic-like as their strangeness fraction drops below 3\%.
\end{abstract}
\begin{keyword}
Equation of state \sep Hyperonic stars\sep Rapid rotation \sep Gravitational wave events
\end{keyword}
\end{frontmatter}
\section{Introduction}
\label{sec:Intro}
Compact stars (CSs) provide unique laboratories to probe dense matter
under extreme conditions which cannot be reproduced on Earth. The
composition of the deep interiors of CS is not known. It represents the
main uncertainty for the determination of the static and dynamic
properties of CS.  Various high-density compositions have been studied
assuming different degrees of freedom, for example, compositions
featuring purely nucleonic, heavy baryon-admixed, and/or deconfined
quarks matter, for reviews see Refs.~\citep{Weber:2004,Sedrakian:2007,Lattimer:2015,Watts:2016,Oertel:2016,Chatterjee:2016,Yagi:2017,Baym:2018,Sedrakian:2019,Baiotti:2019,Lijj:2019b,Llanes-Estrada:2019,Malfatti:2020,Tolos:2020,Sedrakian:2021,Vidana:2021,Burgio:2021}.
In particular, hyperons have been studied as an option as their
nucleation may become energetically favorable above a threshold, which is distinct for each hyperon and is controlled by the conditions of $\beta$-equilibrium and charge neutrality among the baryons and leptons, see Fig.~5 of Ref.~\citep{Lijj:2018a}.
Hyperonization of dense matter then reduces the pressure of
dense matter which has a significant impact on the maximum CS mass, for
reviews see Refs.~\cite{Tolos:2020,Sedrakian:2021,Vidana:2021,Burgio:2021}.

Currently, the most rigorous constraints on the high-density behavior
of the equation of state (EoS) come from the observations of a few
massive pulsars with masses
$\sim 2.0\,M_{\odot}$~\citep{Demorest:2010,Antoniadis:2013,NANOGrav:2019,Fonseca:2021,Romani:2022}.
These observations set a lower bound on the maximum mass of CS
predicted by any model of dense matter.  The long-awaited detection of
gravitational waves (GWs) from a binary neutron star merger, the
GW170817 event placed significant constraints on the tidal
deformability of canonical-mass stars and thus provided additional
constraints on the EoS of dense matter at intermediate
densities~\citep{LVC:2017,LVC:2018,LVC:2019}.  The multi-messenger
analyses of GW170817 event suggest that the maximum mass of static CS
may not exceed $\sim 2.3\,M_{\odot}$~\citep{Margalit:2017,Shibata:2017,Ruiz:2017,Rezzolla:2018,Khadkikar:2021}. The X-ray pulse profile modeling of pulsars with data from the NICER observatories recently led to measurements of CS radii. The estimates
for one of the most massive known pulsar, PSR
J0740+6620~\citep{Riley:2021,Miller:2021}, open prospects of
constraining the properties of the EoS, in particular, the composition of matter at high densities.  PSR J0740+6620 has a mass of about
$\sim 2.1\,M_{\odot}$ and is thus about 50\% more massive than PSR
J0030+0451~\citep{Miller:2019,Riley:2019}, yet current measurements do
not indicate a significant difference in their sizes.  This may
indicate that the turning point, i.e., the maximum, of the mass-radius
relation occurs above the mass of PSR
J0740+6620~\citep{Legred:2021,Huth:2022}.  Previous models of
hyperonic stars were mainly constrained by the masses of massive
pulsars~\citep{Bednarek:2011,Massot:2012,Weissenborn:2012a,Weissenborn:2012b,Colucci:2013,Providencia:2013,vanDalen:2014,Gomes:2014,Lopes:2014,Oertel:2015,Maslov:2015,Tolos:2016,Fortin:2017,Lijj:2018a,Fortin:2020,Stone:2022,Tuzh:2022,Liang:2022}.

Observational identification of neutron stars (black holes) as members
of binary systems requires the knowledge of the upper (lower) limit on
the gravitational mass of a neutron star (black hole). The GW190814
event~\citep{LVC:2020}, caused by the merger of two stellar objects 
with an extremely asymmetric mass ratio, contained a primary black 
hole with a mass of $23.2^{+1.1}_{-1.0}\,M_{\odot}$. The secondary's 
mass was in the range of $2.59^{+0.08}_{-0.09}\,M_{\odot}$. The latter 
value of mass is within the hypothesized ``mass-gap" between neutron 
stars and black holes, $2.5 \lesssim M/M_{\odot} \lesssim 5$, where no 
compact object had ever been observed before. Whether the light 
companion is the most massive neutron star or the lightest black hole discovered so far is unclear
yet~\citep{Most:2020,Sedrakian:2020,Lijj:2020,Nathanail:2021,Fattoyev:2020,Zhangnb:2020,Tsokaros:2020,Huangkx:2020,Biswas:2021,Tews:2021,Dexheimer:2020,Tanhuang:2020,Bombaci:2020,Roupas:2020,Christian:2020,Demircik:2020,Jumin:2021}. 
Recently, the event GW200210~\citep{LVKC:2021} was reported in which
the components have masses of $24.1^{+7.5}_{-4.6}\,M_{\odot}$ and
$2.83^{+0.47}_{-0.42}\,M_{\odot}$.  In
Refs.~\citep{Sedrakian:2020,Lijj:2020} we suggested that the secondary
component of GW190814 is more likely a black hole rather than a CS, by
considering hyperonic EoS models where hyperonic couplings to vector
mesons were based on the SU(6) quark model, while those to
scalar mesons were fitted to the depth of their potential at nuclear
saturation density.

The main motivation of this work is to extend the previous
studies~\citep{Sedrakian:2020,Lijj:2020} of massive hyperonic 
CS from SU(6) symmetry based
vector meson couplings to those that arise within the more general
SU(3) symmetry~\citep{deSwart:1963} which was implemented in the
context of CSs within Hartree~\cite{Schaffner:1994} and Hartree-Fock
\cite{Lijj:2018a} based CDF models. This provides a more complete
exploration of the parameter space that admits the existence of
massive hyperonic stars. Indeed, the SU(6) model combines the flavor
SU(3) and spin SU(2) symmetries, which is a special case of the more
general SU(3) model~\citep{deSwart:1963}.
Previously, several authors explored the effect of breaking of SU(6) 
symmetry down to SU(3) for selected nucleonic EoS in the vector-meson 
sector~\citep{Weissenborn:2012b,Lijj:2018a}, 
scalar-meson sector~\cite{Colucci:2013} and both \cite{Lopes:2014}. 
In the scalar-meson sector the SU(3) relations do not hold after fixing 
the $\Lambda$-hyperon depth~\cite{Fortin:2017,Sedrakian:2020}
and the hyperonic couplings are fixed by the values of hyperonic potential 
depths. Therefore, the SU(3) relations are useful for the vector-meson sector 
only. These works demonstrated that within the SU(3) symmetric models for 
the vector-meson sector it is possible to construct massive hyperonic CSs with
maximum masses as high as 2.2-$2.3\,M_\odot$.  They used models 
with fixed properties of nucleonic component within the relativistic mean 
field models, which preclude by construction a study of the dependence of 
the results on continuous variations of nuclear matter 
characteristics such as symmetry energy, its slope as well as the skewness.

This paper is organized as follows. In Sec.~\ref{sec:frame} we briefly
outline the key features of the CDF model for hyperonic matter.
Section~\ref{sec:result} discusses the bulk properties of hyperonic 
stars predicted by our CDF approach for a broad range of variations of 
the parameters. We discuss the implications of these models for the
interpretation of GWs produced in binary stellar collisions involving 
massive secondaries whose masses lies in the ``mass-gap".  Finally, 
a summary of our results is provided in Sec.~\ref{sec:conclusion}.

\section{CDF model for hypernuclear matter}
\label{sec:frame}
We use in this work the standard CDF theory with density-dependent
meson-baryon couplings for a many-body nuclear system whose interaction
Lagrangian is given by~\cite{Typel:1999,Lalazissis:2005,Lijj:2018a,Lijj:2019a}
\begin{align}\label{eq:interaction_Lagrangian}
\mathscr{L}_{\text{int}}
& = \sum_B \bar{\psi}_B\Big(-g_{\sigma B}\sigma-g_{\sigma^\ast B}\sigma^\ast
    -g_{\omega B}\gamma^\mu\omega_\mu \nonumber \\
& \hspace{1.0em} -g_{\phi B}\gamma^\mu\phi_\mu -g_{\rho B}\gamma^\mu\vec{\rho}_\mu\cdot\vec{\tau}_B\Big)\psi_B,
\end{align}
where $\psi_B$ stands for the Dirac spinors. The index $B$ labels the
$J^P = \frac{1}{2}^+$ baryonic octet with the member masses denoted
by $m_B$.  The explicit form of the free Lagrangian can be found in
Ref.~\cite{Typel:1999,Lalazissis:2005,Lijj:2018a}. The octet of baryons interacts via exchanges
of $\sigma,\,\omega$, and $\rho$ mesons, which comprise the minimal set
necessary for a quantitative description of nuclear
phenomena~\citep{Serot:1997,Vretenar:2005}. We consider further two
hidden-strangeness mesons{\color{red},} $\sigma^\ast$ and $\phi$, which describe
interactions between
hyperons~\cite{Schaffner:1994,Oertel:2015,Lijj:2018a,Raduta:2017}.
The mesons couple to baryons with coupling constants $g_{mB}$, which
are functions of the baryonic density,
$g_{mB} = g_{mB}(\rho_{\rm sat})f_m(r)$, where
$r = \rho/\rho_{\rm sat}$ with $\rho_{\rm sat}$ being the nuclear saturation
density. For the explicit form of the functions $f_m(r)$ see
Refs.~\citep{Lalazissis:2005,Lijj:2018a}. The interaction
Lagrangian~\eqref{eq:interaction_Lagrangian} is fixed by first
assigning the baryons and mesons their masses in the vacuum. Next, one
fixes the three coupling constants
($g_{\sigma N}, g_{\omega N}, g_{\rho N}$) in the nucleonic sector and
the four parameters that enter the functions $f_m(r)$.  Then ground
state properties of infinite nuclear matter and finite nuclei can be
computed uniquely in terms of the above seven adjustable parameters.
Note that the constraint conditions on the $f_m(r)$ function reduce the
eight parameters for $\sigma$ and $\omega$-mesons to three~\citep{Lalazissis:2005,Lijj:2018a}. In addition, there is one parameter for the $\rho$-meson. 
There are in total four parameters that enter the functions $f_m(r)$. 

The EoS of isospin asymmetric nuclear matter can be expanded around 
nuclear saturation and the isospin symmetric limit in
power series~\citep{Margueron:2018,Lijj:2019a}
\begin{align}
\label{eq:Taylor_expansion}
E(n, \delta) & \simeq  E_{\rm{sat}} + \frac{1}{2!}K_{\rm{sat}}n^2 + \frac{1}{3!}Q_{\rm{sat}}n^3 \nonumber \\
             & \hspace{1.0em} + E_{\rm{sym}}\delta^2 + L_{\rm{sym}}\delta^2n + {\mathcal O}(n^4,n^2\delta^2),
\end{align}
where $n =(\rho-\rho_{\rm sat})/3\rho_{\rm sat}$ and
$\delta = (\rho_{\rm n}-\rho_{\rm p})/\rho$. The coefficients of the
density-expansion in the first line of Eq.~\eqref{eq:Taylor_expansion}
are the characteristic coefficients of nuclear matter in the isoscalar
channel, specifically, the saturation energy $E_{\rm sat}$,
incompressibility $K_{\rm sat}$, and skewness $Q_{\rm sat}$. The
coefficients associated with the expansion away from the symmetric
limit in the second line are the characteristic parameters in the
isovector channel, i.e., the symmetry energy $E_{\rm sym}$ and its
slope parameter $L_{\rm sym}$.  The quantities which arise at a higher
order of the expansion, specifically $\Qsat$ and $\Lsym$, are only
weakly constrained by the conventional fitting protocol used in
constructing the density functionals, i.e., the procedure which
involves usually fits to nuclear masses and radii.  However, the value
of $\Qsat$ controls the high-density behavior of the nucleonic energy
density, while the value of $\Lsym$ determines the intermediate-density 
behavior of the nucleonic energy density according to Eq.~\eqref{eq:Taylor_expansion}.

It is interesting to examine the potentials of the baryons in pure
neutron matter, given by
\begin{align}
\hspace{-1.5em}
V_B = - g_{\sigma B}\bar{\sigma} -g_{\sigma^\ast B}\bar{\sigma}^\ast
      + g_{\omega B}\bar{\omega} + g_{\phi B}\bar{\phi} 
      + g_{\rho B}\tau_{3 B}\bar{\rho} + \Sigma_R,
\end{align}
where the meson fields are replaced by their respective expectation
values in the Hartree mean-field approximation~\citep{Typel:1999,Lalazissis:2005,Colucci:2013,Lijj:2018a},
and $\Sigma_R$ denotes the rearrangement term that comes from the
density-dependence of the meson-baryon coupling
constants~\citep{Typel:1999,Lalazissis:2005,Colucci:2013,Lijj:2018a}.

In the SU(3) model three parameters are describing the
deviation from SU(6) flavor-spin symmetry.  Considering the vector
meson sector, the parameter $\alpha_{\rm v}$ is the weight factor for
the contributions of the symmetric and antisymmetric couplings. Its
SU(6) value is $\alpha_{\rm v} = 1$. Another parameter is the mixing
angle ~$\theta_{\rm v} $ which relates the physical mesons to their
pure octet and singlet counterparts. And, finally, the third parameter
$z$ is the ratio of the meson octet and singlet
couplings~\citep{Dover:1985,Rijken:1998,Weissenborn:2012b,Lijj:2018a}.

\begin{figure}[tb]
\centering
\includegraphics[width = 0.45\textwidth]{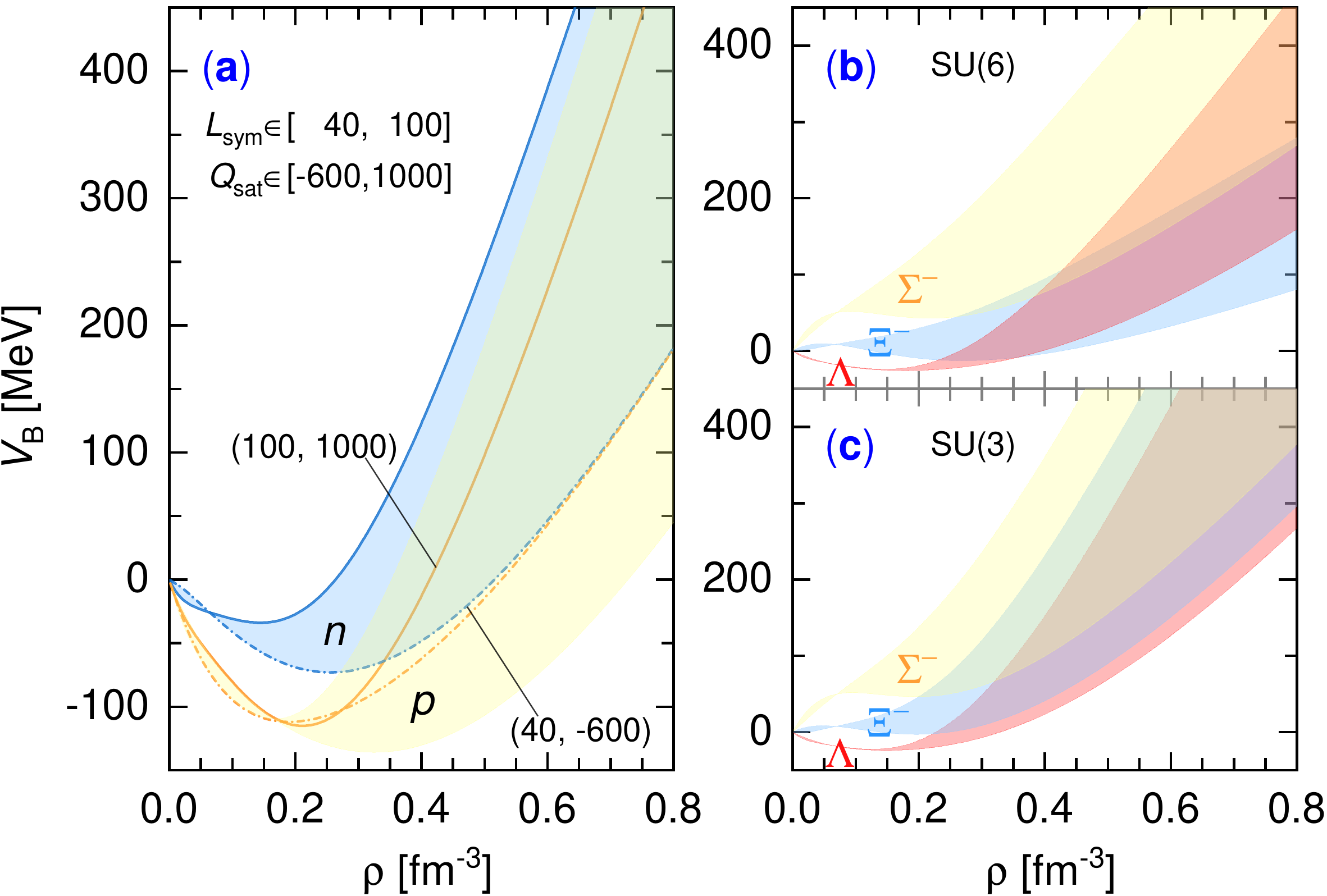}
\caption{ The ranges of single-particle potentials of baryons in
  pure neutron matter as a function of density that are explored in
  this work.  (a) Nucleonic potentials for nucleonic models with
  $\Lsym \in [40, 100]$~MeV and $\Qsat \in [-600, 1000]$~MeV.  The
  results for the stiffest model with
  $(\Lsym, \Qsat) = (100, 1000)$~MeV and the softest one with
  $(40, -600)$~MeV are illustrated explicitly.  (b) Hyperonic
  potentials for the SU(6) model ($z = 1/\sqrt{6}\approx 0.4082$), and (c)
  an extreme case of the SU(3) model ($z =0$). }
\label{fig:potentials}
\end{figure}

The roles played by the parameters $\Qsat$ and $\Lsym$ for the
single-particle potentials of baryons are shown in Fig.~\ref{fig:potentials}, panel (a), where the nucleonic
potentials are shown for models with $\Lsym \in [40, 100]$\,MeV 
and $\Qsat \in [-600, 1000]$\,MeV, and in panels (b) and (c) where
the hyperonic potentials are shown for the cases of SU(6) and extreme SU(3)
with $z=0$.

Given the five macroscopic coefficients in
Eq.~\eqref{eq:Taylor_expansion} together with the preassigned values
of $\rho_{\rm sat}$ and Dirac mass $M^\ast_D$~\citep{Lijj:2018a}, we 
could determine uniquely the seven adjustable parameters of the
Lagrangian~\eqref{eq:interaction_Lagrangian}. In
Ref.~\cite{Lijj:2019a} it has been suggested that one can generate a set of
nucleonic CDF models by varying only one coefficient in
Eq.~\eqref{eq:Taylor_expansion} while keeping the others fixed.
Having this in mind, we map the nucleonic EoS given by the
Lagrangian~\eqref{eq:interaction_Lagrangian} for each set of parameters
$Q_{\rm sat}$ and $L_{\rm sym}$.  For our analysis below we adopt the
lower-order coefficients in Eq.~\eqref{eq:Taylor_expansion}, i.e.,
$E_{\rm sat} = -16.14$, $K_{\rm sat} = 251.15$\,MeV, and
$E_{\rm sym} = 32.31$\,MeV, as those inferred from the DDME2
parametrization~\citep{Lalazissis:2005,Lijj:2019a}, which was
adjusted to the properties of finite nuclei.

The determination of the meson-hyperon couplings $g_{mY}$ represents a
long-standing theoretical challenge due to the lack of sufficiently
abundant and accurate experimental data.  In the present work, we
restrict our attention only to the three lightest quark flavors and
adopt the flavor SU(3) symmetric
model~\citep{deSwart:1963,Schaffner:1994,Lijj:2018a}. To explore the
parameter space associate with the SU(3) model we proceed by assuming
an ideal mixing value of $\theta_{\rm v} = \rm tan^{-1} (1/\sqrt{2})$~\citep{ParticleDataGroup:2020}.
This fixation of the mixing angle $\theta$, which describes the
mixing between the singlet and the octet members of a physical
isoscalar vector mesons, is motivated by the fact that the mixing
between nonstrange and strange quark wave functions in the $\omega$
and $\phi$-mesons is ideal, i.e., the mixing angle assumes the ideal mixing
value quoted above. In addition, from the quadratic mass
formula for mesons, one obtains $\theta \approx 40^\circ$~\citep{ParticleDataGroup:2020}, 
a value that is a very close to the ideal mixing angle 
$\theta \approx 35.3^\circ$. Thus, it is reasonable to keep the 
condition of ``ideal mixing" for the isoscalar vector mesons.  
The dependence on the remaining parameters, $\alpha_{\rm v}$ and $z$, can 
be explored by fixing one of them and varying the other. This has been 
done previously in Ref.~\citep{Lijj:2018a} (see their Figs.~11 and 12) 
showing that reducing the value of either $\alpha_{\rm v}$ or $z$ from their SU(6)
values at fixed value of the other parameter  
yields qualitatively similar modifications of the EoS and the
particle fractions.
We thus choose to vary only one of them, i.e., $z$, while
keeping $\alpha_{\rm v}=1$ fixed at its SU(6) value.

Then we are left with a single free parameter $z$ to quantify
the effects of the SU(3) symmetric model.  In this case, the hyperonic
coupling constants are defined as~\citep{Weissenborn:2012b,Lijj:2018a} 
\begin{subequations}\label{eq:SU3Z}
\begin{align}
\frac{g_{\omega \Lambda}}{g_{\omega N}} =  \frac{g_{\omega \Sigma}}{g_{\omega N}}
&= + \frac{\sqrt{2}}{\sqrt{2} + \sqrt{3}z} \simeq 1-\sqrt{\frac{3}{2}}z,\\
\frac{g_{\omega \Xi}}{g_{\omega N}}
&= + \frac{\sqrt{2} - \sqrt{3}z}{\sqrt{2} + \sqrt{3}z}
  \simeq 1-\sqrt{6}z,\\
\frac{g_{\phi \Lambda}}{g_{\omega N}} =  \frac{g_{\phi \Sigma}}{g_{\omega N}}
&= -\frac{1}{\sqrt{2} + \sqrt{3}z} \simeq -\frac{1}{\sqrt{2}}+ \frac{\sqrt 3}{2} z, \\
\frac{g_{\phi \Xi}}{g_{\omega N}}
&= -\frac{1 + \sqrt{6}z}{\sqrt{2} + \sqrt{3}z}
    \simeq -\frac{1}{\sqrt{2}} - \frac{\sqrt 3}{2} z,
\end{align}
\end{subequations}
where in each equation the last relation shows the $z\to 0$
asymptotes neglecting terms $\mathcal{O}(z^2)$. These asymptotic values
can be compared with the SU(6) values of the coupling constants,
\begin{subequations}
\begin{align}
\frac{g_{\omega \Lambda}}{g_{\omega N}} = \frac{g_{\omega \Sigma}}{g_{\omega N}}
&= \frac{2}{3}, \quad \hspace{1.52em}
\frac{g_{\omega \Xi}}{g_{\omega N}} 
=\frac{1}{3},\\
\frac{g_{\phi \Lambda}}{g_{\omega N}} = \frac{g_{\phi \Sigma}}{g_{\omega N}}
&=-\frac{\sqrt{2}}{3}, \quad
\frac{g_{\phi \Xi}}{g_{\omega N}}
 =-\frac{2\sqrt{2}}{3}.
\end{align}
\end{subequations}
It is seen that the $z=0$ limit of the SU(3) model implies a much stronger repulsive 
interaction among hyperons due to $\omega$-exchange.  

In the SU(6) symmetric model, the $\phi$-meson has a vanishing $\phi$-$N$ coupling, 
whereas it does couple to the nucleon in SU(3) symmetric model in terms of 
\begin{align}\label{eq:SU3N}
\frac{g_{\phi N}}{g_{\omega N}}
= \frac{\sqrt{6}z - 1}{\sqrt{2} + \sqrt{3}z} \simeq
-\frac{1}{\sqrt{2}} +\frac{3\sqrt 3}{2}z,
\end{align}
where the last relation is the $z\to 0$ asymptote as above.

To ensure this new coupling scheme does not spoil the fits to the purely nuclear data, 
we make the replacement~\citep{Lijj:2018a}
\begin{align}
\frac{\tilde{g}^2_{\omega N}}{m^2_\omega}
=  \frac{g^2_{\omega N}}{m^2_\omega} + \frac{g^2_{\phi N}}{m^2_\phi},
\end{align}
where the $\tilde{g}_{\omega N}$ denotes the coupling for the case of
$g_{\phi N} = 0$.  For such a scheme, 
it has been shown in Ref.~\citep{Lijj:2018a} that the EoS of purely nucleonic matter is (almost)
independent of the appearance of $\phi$-meson. For the isovector meson
$\rho$, one has~\citep{Weissenborn:2012b,Lijj:2018a}
\begin{align}\label{eq:SU3Zr}
\frac{g_{\rho \Lambda}}{g_{\rho N}} = 0, \quad
\frac{g_{\rho \Sigma}}{g_{\rho N}} = 2, \quad
\frac{g_{\rho \Xi}}{g_{\rho N}} = 1.
\end{align}

The isoscalar-scalar meson-hyperon couplings
are then determined by fitting them to certain preselected properties of
hypernuclear systems. We fix the coupling constants $g_{\sigma Y}$
using the following hyperon potentials in symmetric nucleonic matter
at saturation density, $\rho_{\rm sat}$, extracted from hypernuclear phenomena~\citep{Feliciello:2015,Gala:2016}:
\begin{align}
\label{eq:hyp_potentials}
U^{(N)}_\Lambda = -U^{(N)}_\Sigma= -30~\text{MeV},\quad U^{(N)}_\Xi = -14~\text{MeV}.
\end{align}
Finally, we use the estimate
$U^{(\Lambda)}_\Lambda (\rho_{\rm{sat}}/5) = - 0.67~\text{MeV}$,
which is extracted from the $\Lambda\Lambda$  bond
energy~\citep{KEK:2013}, to fix the coupling constant
$g_{\sigma^\ast \Lambda}$. The couplings of the remaining hyperons $\Xi$
and $\Sigma$ to the $\sigma^\ast$-meson are determined by the relation
$g_{\sigma^\ast Y}/g_{\phi Y} =
g_{\sigma^\ast\Lambda}/g_{\phi\Lambda}$.  In this manner, we assume
that the hyperon potentials scale with density as the nucleonic
potential, therefore their high-density behavior is inferred from that
of the nucleons.  In Fig.~\ref{fig:potentials}\,(b) we show the
potentials of hyperons in pure neutron matter for two limiting cases,
$z = 1/\sqrt{6}$ and 0. The former corresponds to the SU(6) model
while the latter is the extreme case of the SU(3) model.  It is seen that
the hyperon potential depths~\eqref{eq:hyp_potentials}, indeed,
determine only the EoS region around the saturation density.  In
contrast, the meson coupling constants~\eqref{eq:SU3Z} affect largely
the high-density regime of the EoS and consequently the degree of
stiffness of the EoS, which is closely linked to the inner core composition of CSs.

For any set of coupling constants 
the EoS of the core of a CS is determined 
by applying the conditions of weak equilibrium 
and change neutrality. This EoS is then 
matched (interpolated) smooth\-ly to the  EoS of the crustal matter  
given by Refs.~\citep{Baym:1971a,Baym:1971b} at 
the core-crust transition 
density $\sim \rho_{\rm sat}/2$. The details of 
core-crust matching procedure and the model of the crust EoS 
affect to some extent 
the value of the radius and, therefore, the tidal deformability
for light CSs~\citep{Fortin:2016,Piekarewicz:2019}, but 
the uncertainties are negligible for massive stars of interest herein.

\section{Gross properties of hyperonic stars}
\label{sec:result}
In this section, we will explore the gross parameters of hyperonic
stars for a selected set of parameters that control the stiffness of
the EoS.  In the nucleonic sector we vary the characteristics
of nuclear matter $\Qsat$ and $\Lsym$. In the hyperonic sector,
we vary the parameter $z$ associated with the breaking of the SU(3)
symmetry.  Our choice of parameters that describe the EoS of 
hypernuclear matter is as follows:
\begin{itemize}
\item (I) Soft EoS in the nucleonic sector with $\Lsym = 40$\,MeV,
  which is close to the lower values of the 90\% confidence ranges of the
  PREX-2 neutron skin
  measurement~\citep{PREX-II:2021,Reed:2021,Reinhard:2021}. Our EoS
  predicts for $1.4\, M_{\odot}$ mass star a radius and tidal
  deformability in the ranges of $11.8 \lesssim R_{1.4} \lesssim 13.2$~km
  and $280 \lesssim \Lambda_{1.4} \lesssim 750$ when the remaining
  parameters $\Qsat$ and $z$ are varied.  These values are
  within the range derived for the multimessenger GW170817
  event~\citep{LVC:2018,LVC:2019}.
\item (II) Stiff EoS in the nucleonic sector with $\Lsym = 100$\,MeV,
  which is close to the central value of the PREX-2
  analysis~\citep{Reed:2021}.  In this case, we find that the radius
  and deformability are larger, their values for a $1.4M_{\odot}$
  mass star being in the range of $12.8 \lesssim R_{1.4} \lesssim 14.3$~km 
  and $450 \lesssim \Lambda_{1.4} \lesssim 1200$. These values are in
  agreement with the mass and radius inferences from the NICER experiment
  for PSR J0030+0451~\citep{Miller:2019,Riley:2019}, but are outside
  of the range deduced from the GW170817 event~\citep{LVC:2018,LVC:2019}.
  Exceptions to this are models with $\Qsat \lesssim -400$\,MeV.
\end{itemize}
%
\begin{figure}[tb]
\centering
\includegraphics[width = 0.45\textwidth]{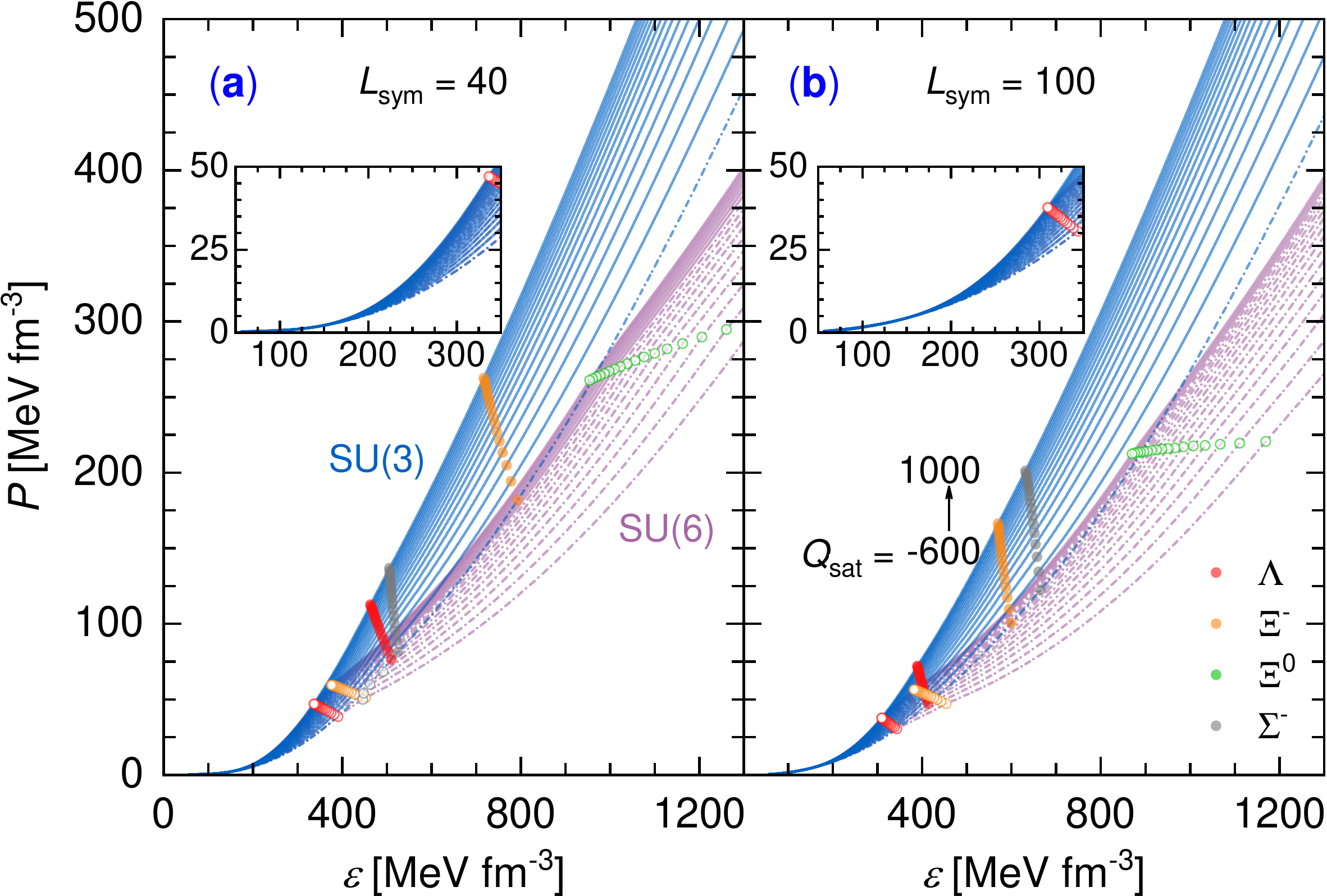}
\caption{ The EoS for SU(6) and SU(3) symmetric models with
  $z = 1/\sqrt{6}$ and $z=0$, respectively. The intermediate density
  nucleonic EoS is either soft [panel (a)] or stiff [panel (b)] depending on
  the values of $\Lsym = 40$ and 100\,MeV. The stiffness of the
  high-density nucleonic component is explored by varying $\Qsat$ in
  the range $[-600,1000]$ MeV with a step size of 100 MeV.  Hyperonic EoS that
  produce stars with $M^{\rm max}_{\rm TOV} \geq 2.0\,M_\odot$ are
  shown by solid lines, those with
  $M^{\rm max}_{\rm TOV} < 2.0\,M_\odot$ are shown by dashed lines. The
  onsets of hyperons are marked by open circles for the SU(6) symmetric
  model and by filled circles for the SU(3) symmetric model. }
\label{fig:EoS}
\end{figure}

Figure~\ref{fig:EoS} shows a collection of EoS that cover the
relevant range of parameters both for the nucleonic and hyperonic
sectors. The model EoS is distinguished by (a) the values of
$\Lsym=40$ and 100\,MeV which control the intermediate-density
stiffness in the nucleonic sector; (b) the values of $\Qsat$ which
control the high-density stiffness of the nucleonic sector and are
drawn from the interval $[-600, 1000]$~MeV with a step size of 100~MeV; (c) the
values of the $z$-parameter which takes on two values:
$z = 1/\sqrt{6}$ for the SU(6) model and $z =0$, which is an extreme
case of the SU(3) model. As can be seen, the intermediate-density soft
models show a delay in the appearance of hyperons as the density is
increased.  As a consequence, the EoS is stiffer at high
densities once the hyperons are admixed with the nucleonic matter.
Note the different ordering of the thresholds of the appearance of
hyperons in the SU(3) and SU(6) models. In the SU(6) case, $\Lambda$
hyperons appear first and are followed by $\Xi^-$, then $\Xi^0$
hyperons as the density is increased. In the SU(3) model, $\Lambda$'s
are followed by the $\Sigma^-$ hyperons and the onset of $\Xi^0$ is
shifted to densities that are not relevant for stable CSs.  Note
that here and below we will keep occasionally the EoS models with
maximum masses below the $2.0\,M_{\odot}$ mass limit to account for the
possibility of two families of CSs, in which case
stars with masses of $2.0\,M_{\odot}$ and higher
are strange stars~\cite{Drago:2014}.

\subsection{Static sequences of hyperonic stars}
\label{sec:static}
\begin{figure}[tb]
\centering
\includegraphics[width = 0.45\textwidth]{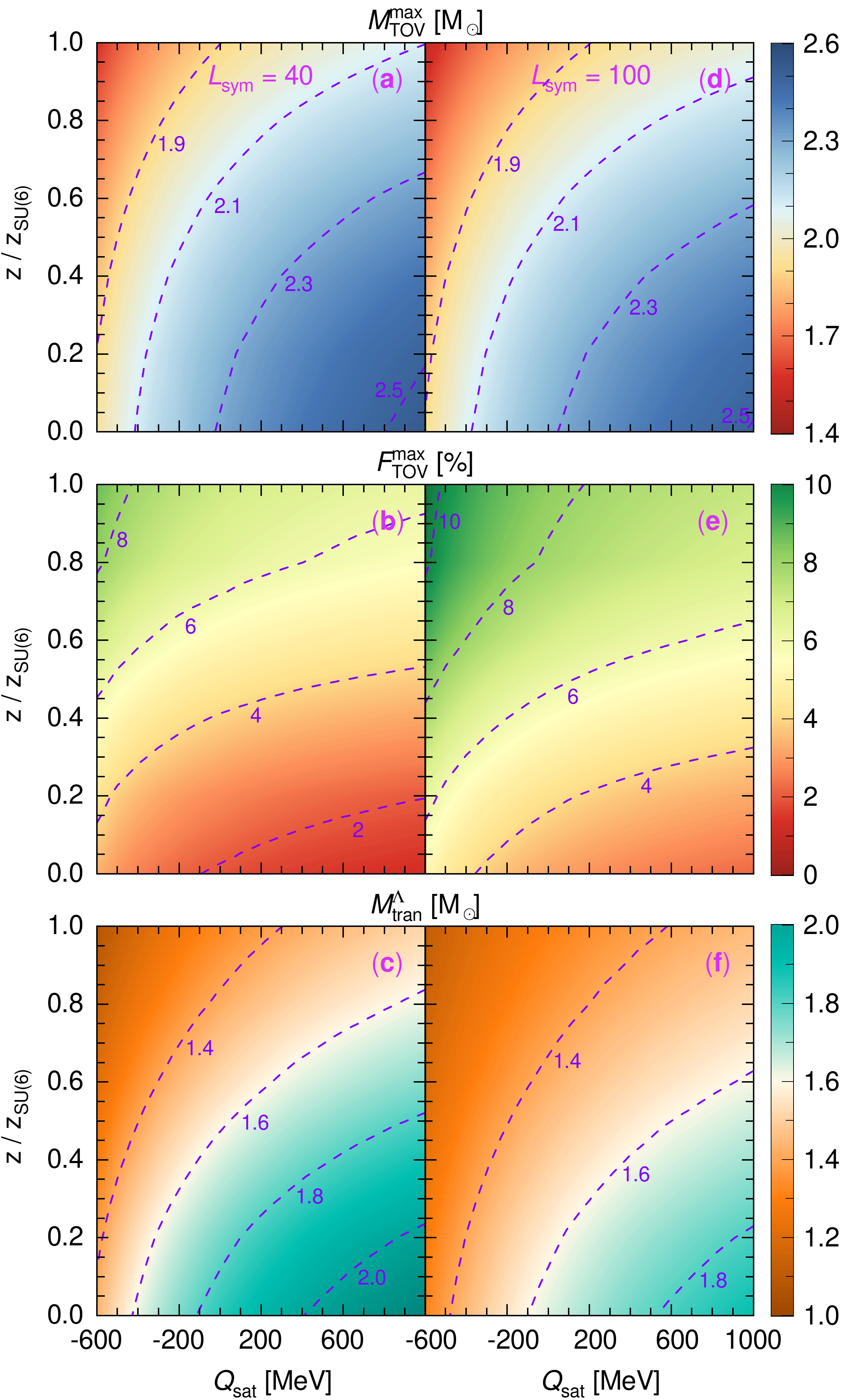}
\caption{ The masses $M^{\rm max}_{\rm TOV}$ (a, d), strangeness fractions $F^{\rm max}_{\rm TOV}$ (b, e) of maximum-mass configurations and the hyperon threshold masses $M^\Lambda_{\rm tran}$ (c, f) of static sequences for hyperonic models within a range of values spanned by $\Qsat$ and $z$. The values for $z$ are normalized by its SU(6) case $z_{\rm {SU(6)}} = 1/\sqrt{6} \approx 0.4082$. The left and right panels show, respectively, results for models with $\Lsym = 40$ and 100~MeV in the nucleonic sector. }
\label{fig:MF_static}
\end{figure}

We start by considering sequences of static (non-rotating) stars which
are described by the Tolman-Oppen\-heimer-Volkoff (TOV)
equation for a given input EoS.  Figure~\ref{fig:MF_static} shows the
maximum mass $M^{\rm max}_{\rm TOV}$, its strangeness fraction
$F^{\rm max}_{\rm TOV}$ ($F \equiv N_{\rm S}/ N_{\rm B}$ with $N_{\rm S(B)}$ being the total strangeness(baryon) numbers in a star~\cite{Fortin:2020}), and the mass
$M^\Lambda_{\rm tran}$ of the star at which the $\Lambda$ hyperon
first appears, as functions of the $\Qsat$ and $z$ parameters for the two
classes of models with $\Lsym = 40$ and 100\,MeV, as described
above. The parameter ranges are $0\le z\le 1/\sqrt{6}\equiv z_{\rm SU(6)}$ and
$-600\le \Qsat\le 1000$, where the upper value of $z$ corresponds to
its SU(6) value.  

According to the results shown in Fig.~\ref{fig:MF_static} 
the following conclusions can be drawn: 

(i) The upper left corner of the parameter space (low $\Qsat$ and
$z \le z_{\rm SU(6)}$) is inconsistent with the mass measurement of
PSR J0740+6620~\citep{NANOGrav:2019,Fonseca:2021}, i.e., the
consistency of the SU(6) symmetric model requires large values of
$\Qsat$. Moving away from SU(6) symmetry stiffens the EoS and
consequently relaxes the large $\Qsat$ requirement. For example, in the
extreme limit where $z\to 0$ the mass constraint above is met for any value
of $\Qsat$. Models with smaller values of $\Lsym$ (c.f. panels a and d)
predict (counterintuitively) a wider range of parameters that produce massive enough stars,
because smaller $\Lsym$ implies softer nucleonic EoS at the intermediate densities and, therefore, 
{\it delayed onset of hyperons}. 
The strangeness fraction of maximum-mass configurations is
anti-correlated with the maximum masses of stars, since the more
massive the star the smaller the fraction of hyperons and the
strangeness fraction $F^{\rm max}_{\rm TOV}$.  Thus, going away from
the SU(6) symmetry limit suppresses the emergence of hyperons in
massive stars by large factors of $\sim 3$-4. The most massive models
then have a negligible hyperonic content and are close in their
properties to their purely nucleonic stars. According to the lower panels 
of Fig.~\ref{fig:MF_static}, the masses of stars in which the threshold 
for the appearance of $\Lambda$ hyperons is reached shifts to higher values as one moves away from SU(6) $z$ value and increases the value of $\Qsat$. 
This is a direct consequence of the stiffening of the EoS in the
nucleonic sector by larger values of $\Qsat$ and in the hyperonic
sector by smaller values of $z$.  

(ii) A combination of numerical simulations with simple but
  reasonable assumptions lead to the conclusion that the GW170817 event
  resulted in a rapidly rotating neutron star which collapsed to a
  black hole, a scenario that allowed scientists to 
  deduce an approximate upper
  limit for the TOV stellar mass  in the range of
  $2.1\le M^{\rm max}_{\rm TOV}/M_{\odot}\le
  2.3$~\citep{Ruiz:2017,Shibata:2017,Rezzolla:2018,Khadkikar:2021}. The
  upper limit of this value range is obtained if finite
  temperature EoS effects are included in the
  analysis~\citep{Khadkikar:2021}. According to
  Fig.~\ref{fig:MF_static}, the stars in the high-$\Qsat$ and low-$z$
  domain have masses that violate this upper limit. These are
  also the stars with strongly reduced hyperon fractions of 
  $F^{\rm max}_{\rm TOV} \sim 2$-$6\%$ since the
  onset of hyperons occurs  only in  the most massive ($M \gtrsim 1.8\,M_\odot$)
  stars. 

(iii) Finally, note that this high-$\Qsat$ and low-$z$ domain
  features stars with masses $M^{\rm max}_{\rm TOV} \le 2.5\,M_\odot$. 
  Thus the stars from this domain would be compatible with the
  mass of the secondary in the GW190814 event~\citep{LVC:2020} and its
  interpretation as a low-spin star with a trace of hyperons
  ($F^{\rm max}_{\rm TOV} \sim 1$-$2\%$).  Inverting the argument and
  assuming that GW190814 event contained a massive CS one
  can put limits on the values of parameters of the CDF, specifically
  in our case we require $\Qsat > 800$~MeV and
  $z/z_{\rm SU(6)} \lesssim 0.1$.  We recall that the latter
  constraint implies $g_{\omega N} \approx g_{\omega Y}$ and
  $g_{\phi N} \approx g_{\phi Y}$, see Eq.~\eqref{eq:SU3Z}.  We stress
  again that for these models hyperon populations is very small since
  $F^{\rm max}_{\rm TOV} \sim 1\%$. We also note that the requirement
  $\Qsat > 800$~MeV is consistent with the recent nuclear 
  CDFs~\cite{Taninah:2020,Weib:2020} that were calibrated by finite
  nuclei.

\begin{figure}[tb]
\centering
\includegraphics[width = 0.47\textwidth]{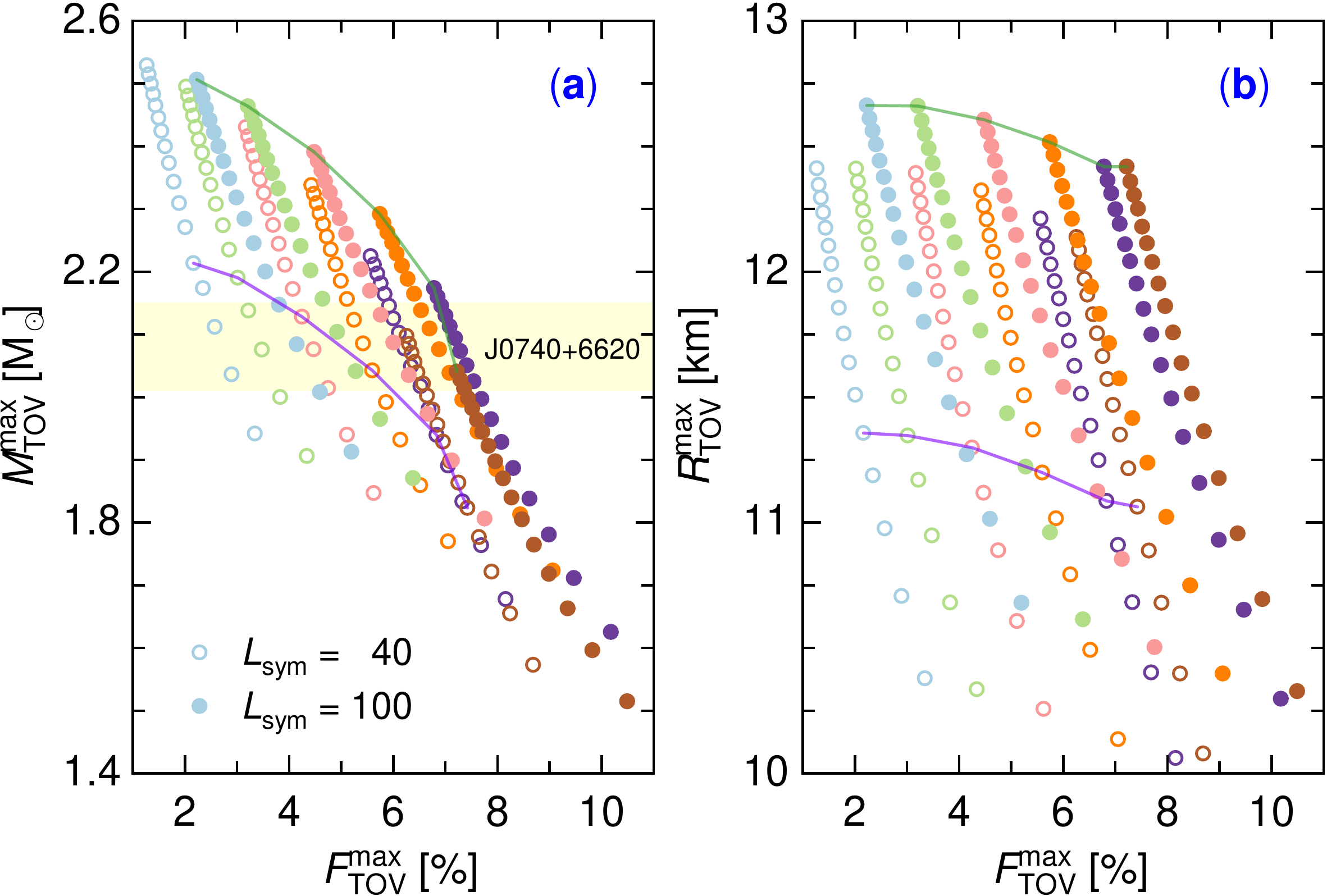}
\caption{ Gravitational mass $M^{\rm max}_{\rm TOV}$ and radius $R^{\rm max}_{\rm TOV}$
  as functions of strangeness fraction $F^{\rm max}_{\rm TOV}$ for
  the maximum-mass configuration calculated for our collection of
  EoS. The open circles denote models with $\Lsym = 40$~MeV, 
  while filled circles refer to those with $\Lsym = 100$~MeV.  The different
  $z$ models are computed for $0\le z\le z_{\rm SU(6)}$ 
  with a step size of $0.2\,z_{\rm SU(6)}$ and are 
  distinguished by different colors.  The same color symbols represent models with fixed $z$ and $\Qsat$ values
  varying in the interval [-600, 1000] with a step size of 100\,MeV.  
  The lines link the models with the same nuclear matter parameters 
  $\Qsat$ and $\Lsym$. The yellow shading indicates the mass of PSR
  J0740+6620~\citep{NANOGrav:2019}. }
\label{fig:MR_static}
\end{figure}

We plot in Fig.~\ref{fig:MR_static} the mass and radius of the
maximum-mass configuration as functions of strangeness fraction in the
core for a collection of our EoS.  As seen in Fig.~\ref{fig:MR_static},
for fixed values of $z$ the mass $M^{\rm max}_{\rm TOV}$ depends linearly on
$F^{\rm max}_{\rm TOV}$.  In the case where $\Qsat$ is fixed, the
relation has a polynomial form. The same scalings also apply to the
radius $R^{\rm max}_{\rm TOV}$ (see also the discussion of the last
relation in Ref.~\citep{Weissenborn:2012b}).  However, the mass or the
radius as a function of the parameters $z$ and $\Qsat$ are randomly
distributed and cannot be easily fitted. Some of the
models shown in Fig.~\ref{fig:MR_static} have maximum masses below the
mass band of PSR J0740+6620~\citep{NANOGrav:2019} and therefore are
ruled out. This conclusion works only within the single CS
family scenario and can be circumvented in a scenario where there is
a separate family of strange stars~\citep{Drago:2014}. In this
scenario very compact hyperonic stars with masses far below
$2.0\,M_{\odot}$ are possible.

\subsection{Keplerian sequences of hyperonic stars}
\label{sec:Keplerian}
\begin{figure}[tb]
\centering
\includegraphics[width = 0.45\textwidth]{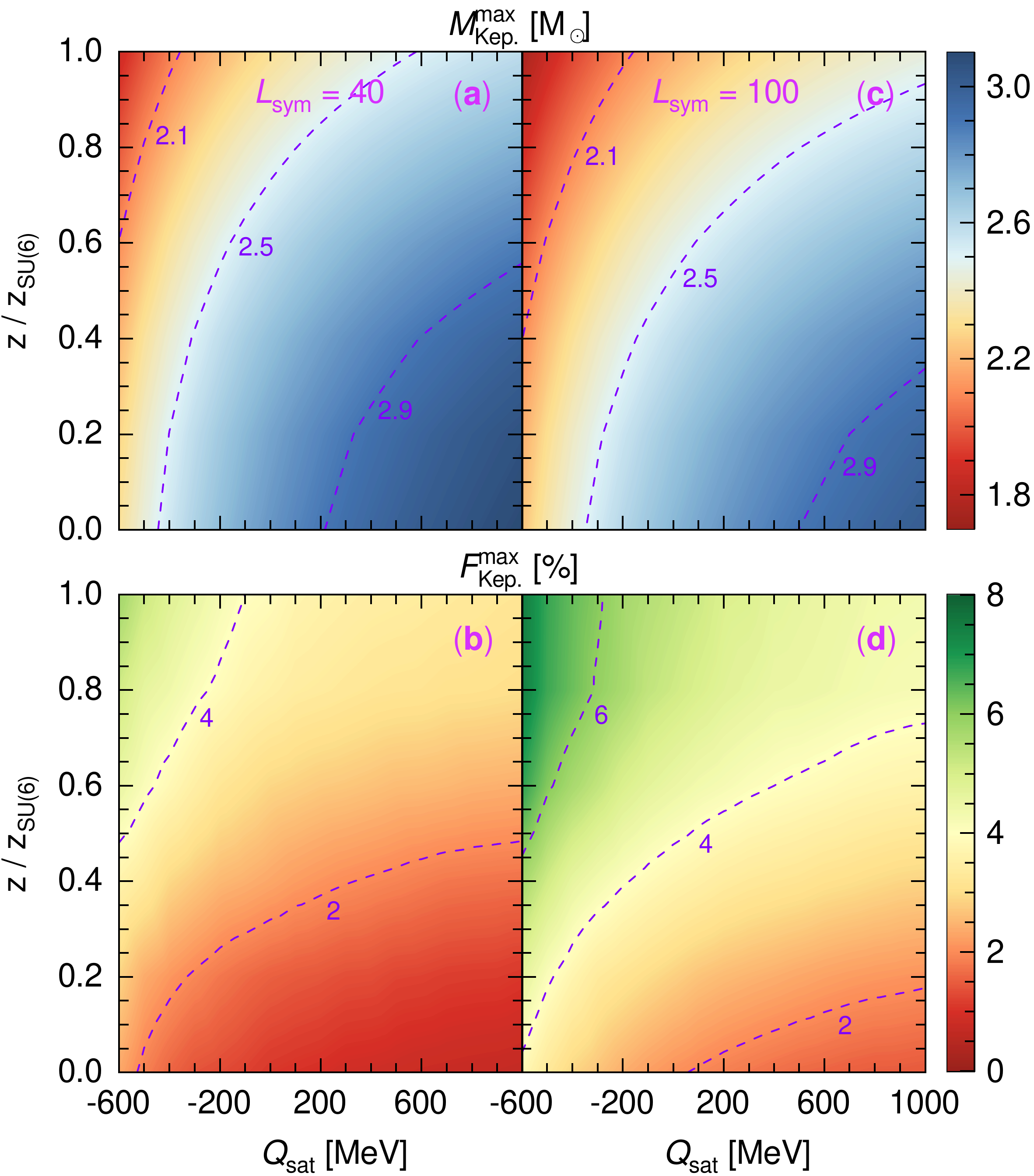}
\caption{ The masses $M^{\rm max}_{\rm Kep.}$ (a, c) and strangeness
  fractions $F^{\rm max}_{\rm Kep.}$ (b, d) of maximum-mass
  configurations of Keplerian hyperonic models for a
  range of values of $\Qsat$ and $z$. The left and right
  panels show, respectively, results for models with $\Lsym = 40$ and
  100~MeV. }
\label{fig:MF_kepler}
\end{figure}

Next we consider uniformly rotating stellar models assuming
stationary, ideal fluid equilibria described by general
relativity~\citep{Weber:1992,Cook:1994,Stergioulas:1994,Haensel:1995}.
Let us first focus on the Keplerian limit of equilibria rotating at
the maximal value of the rotation frequency. Because these configurations
have the maximal value of the centrifugal force they carry the maximum
mass allowed by uniform rotation. The rotating equilibria were
computed with the public domain RNS code~\cite{RNScode}.

The maximum mass of the Keplerian sequence $M^{\rm max}_{\rm Kep.}$
and the corresponding strangeness fraction $F^{\rm max}_{\rm Kep.}$ are
shown in Fig.~\ref{fig:MF_kepler} for varying values of the parameters
$\Qsat$ and $z$ for two classes of models distinguished by the value
of $\Lsym$.  The maximum mass is shifted to higher values compared to
its non-rotating limit by about 20\%, as
expected~\citep{Weber:1992,Cook:1994,Paschalidis:2017}.  In the
large-positive-$\Qsat$ and small-$z$ domain the masses increase up to
values of around $3.0\,M_{\odot}$, which are within the ``mass-gap"
between the measured masses of CSs and black holes.  If the secondary
in the GW190814 event was a rapidly spinning CS, then the
tension between such an interpretation and the underlying EoS models is
resolved. Turning to the strangeness fraction of these massive
objects, we note that for stars with $M \simeq 2.5\,M_\odot$ it is in
the range $F^{\rm max}_{\rm Kep.} \sim 3$-$5\%$, whereas for the 
stars with $M = 3.0\,M_\odot$ we find $F^{\rm max}_{\rm Kep.} \sim 1\%$,
i.e., the hyperons have essentially disappeared.  We conclude that achieving 
large masses in the range $M/M_\odot\sim 2.5$-3.0 requires a significant 
suppression of the hyperon population which can occur for the SU(3) 
model in the limit $z\to 0$. It follows from the discussion above that 
only nucleonic stars (ignoring, for all practical
purposes, the vanishing small amount of hyperons) with a rather stiff EoS 
(large-positive-$\Qsat$ values) can achieve large enough masses
which enter the ``mass-gap'' region. 

\begin{figure}[tb]
\centering
\includegraphics[width = 0.45\textwidth]{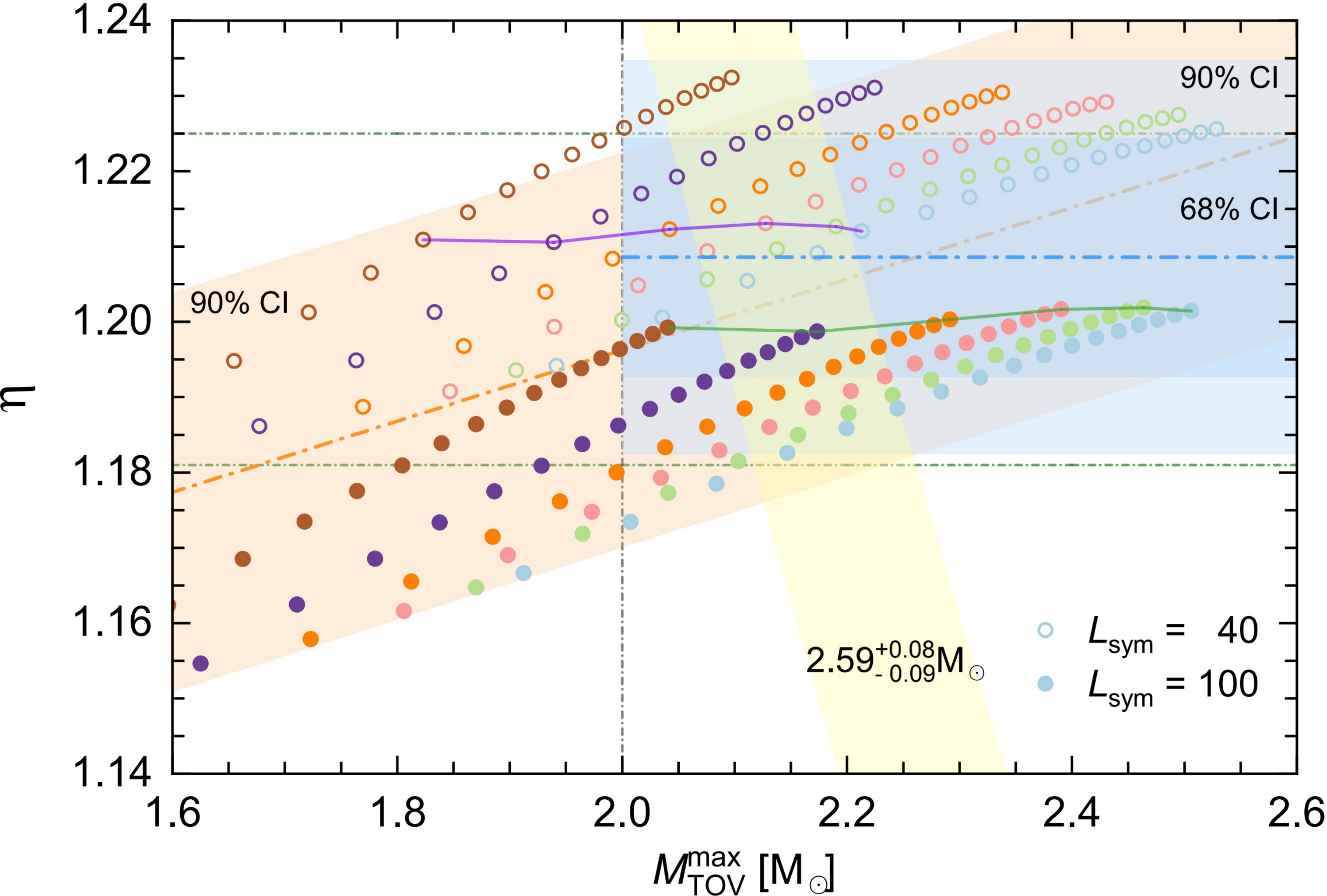}
\caption{ The ratio of
  $\eta=M^{\rm max}_{\rm Kep.}/M^{\rm max}_{\rm TOV}$ as a function of
  $M^{\rm max}_{\rm TOV}$ for a collection of EoS. The open
    circles denote models with $\Lsym = 40$~MeV, while filled circles
    refer to those with $\Lsym = 100$~MeV.  The different $z$ models
    are computed for $0\le z\le z_{\rm SU(6)}$ with a step size of
    $0.2\,z_{\rm SU(6)}$ and are distinguished by different colors.
    The same color symbols represent models with fixed $z$ and $\Qsat$
    values varying in the interval [-600, 1000] with a step size of
    100\,MeV. The lines link the models with the same nuclear
  matter parameters $\Qsat$ and $\Lsym$. The orange and blue bands
  show, respectively, a linear fit
  ($\eta = 0.0472\,M^{\rm max}_{\rm TOV}/M_{\odot} + 1.1018)$ from our
  full data and a constant fit ($\eta = 1.209^{+0.026}_{-0.026}$) from
  data with $M^{\rm max}_{\rm TOV} \geqslant 2.0\,M_\odot$ at 90\%
  CIs.  The horizontal lines correspond to the two limits of
  $\eta=1.203^{+0.022}_{-0.022}$ from Ref.~\citep{Breu:2016}. The
  yellow band denotes the mass range
  $M = 2.59^{+0.08}_{-0.09}\,M_\odot$ (at 90\% CI) for the secondary
  of the GW190814 event~\citep{LVC:2020}. }
\label{fig:MM_kepler}
\end{figure}

Figure~\ref{fig:MM_kepler} shows the mass ratio
$\eta=M^{\rm max}_{\rm Kep.}/M^{\rm max}_{\rm TOV}$ as a function of
$M^{\rm max}_{\rm TOV}$. For many EoS models, this quantity is a
constant. However, it is evident from the figure that the mass ratio
$\eta$ increases with the increase of $M^{\rm max}_{\rm TOV}$.
Furthermore, the smaller the value of $\Lsym$, i.e., the softer
the intermediate density EoS, the larger the value of
$\eta$. Our values of $\eta$ can be compared with those obtained 
from the fits to a large collection of nucleonic EoS~\citep{Breu:2016}, 
which gives $\eta =1.203^{+0.022}_{-0.022}$.
The ratios $\eta$ obtained from models with $\Lsym = 40$\,MeV
and $M^{\rm max}_{\rm TOV} \gtrsim 2.0\,M_\odot$ are
slightly shifted upward with respect to the fit obtained from
nucleonic models, while $\eta$ obtained from models with
$\Lsym = 100$\,MeV and $M^{\rm max}_{\rm TOV} \lesssim 2.0\,M_\odot$
are shifted downward.  

If we consider only the scenario of one family of CSs and require
$M^{\rm max}_{\rm TOV} \gtrsim 2.0\,M_\odot$, a constant value for
$\eta$ of $1.209^{+0.026}_{-0.026}$ (at 90\% CI) is obtained.
If we further assume that the secondary object in the GW190814 event was a rapidly
spinning star rotating at its 
Kepler frequency with a mass in the
range  $M_2 = 2.59^{+0.08}_{-0.09}\,M_\odot$ (at 90\%
CI)~\citep{LVC:2020}, then by using the values of $\eta$
shown in Fig.~\ref{fig:MM_kepler} we can evaluate the possible values
of $M^{\rm max}_{\rm TOV}$ as: $2.15^{+0.11}_{-0.12}\,M_\odot$ (at
90\% CI).

\subsection{GW sources with \ensuremath{M_2 \lesssim 3 M_\odot}
    interpreted as fast rotating compact stars} 
\label{sec:GWsource}
\begin{figure}[tb]
\centering
\includegraphics[width = 0.45\textwidth]{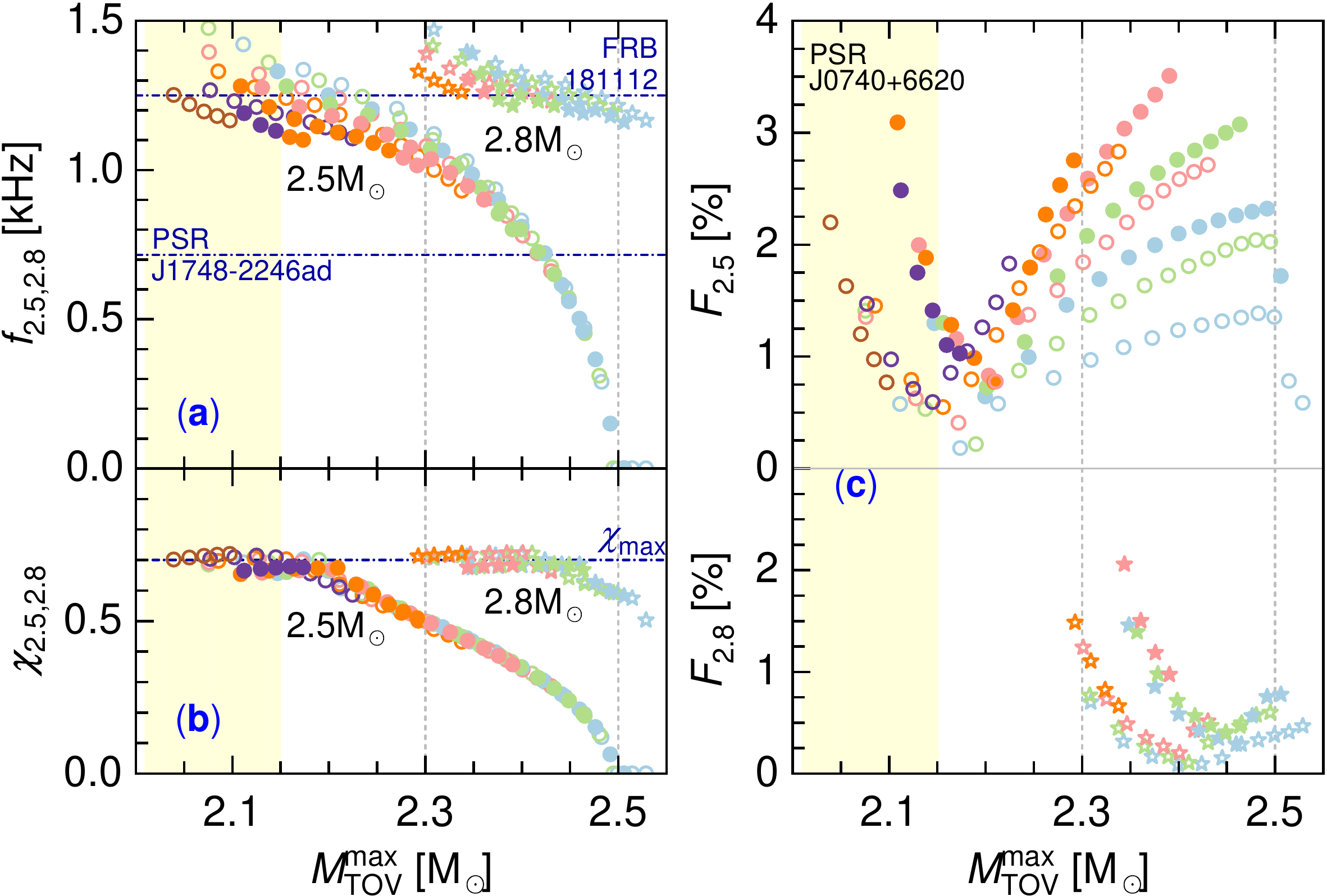}
\caption{ The minimum frequencies $f_{2.5, 2.8}$ (a), dimensionless
  spin parameters $\chi_{2.5, 2.8}$ (b), and strangeness fraction
  $F_{2.5, 2.8}$ (c) of the models which support masses
  $M = 2.5, 2.8\,M_{\odot}$ as a function of the
  $M^{\rm max}_{\rm TOV}$. In panel (a) the lower horizontal line
  corresponds to the frequency of PSR J1748-2246ad~\citep{Hessels:2006}
  and the upper one to that of FRB 181112~\citep{Yamasaki:2020}. In
  panel (b) the horizontal line denotes the upper bound on the spin
  parameter $\chi_{\rm max} = 0.7$ deduced in
  Refs.~\citep{Cook:1994,Haensel:1995}. }
\label{fig:FFM_Rot}
\end{figure}

So far we have generated massive hyperonic stars, both static and fast
spinning, with masses that cover well the range of inferred secondary
masses in GW190814 and GW200210 events \citep{LVC:2020,LVKC:2021} for
which it was deduced that $M_2 = 2.83^{+0.47}_{-0.42}\,M_{\odot}$ and
$M_2 < 3\,M_{\odot}$ with 76\% probability~\citep{LVKC:2021}. At this
point, let us evaluate in addition the minimal frequencies $f_{2.5}$
and $f_{2.8}$ that are necessary to rotationally support stars with
masses of 2.5 and $2.8\,M_{\odot}$, for any given EoS.
These mass-values constitute the lower limit of the 90\% CI
interval for the mass of the secondary in GW190814 and the central
value of the secondary in GW200210, respectively.

Figure~\ref{fig:FFM_Rot}\,(a) shows these frequencies calculated for
our EoS models which predict stars with masses of 2.5 or
$2.8\,M_{\odot}$ either in the static limit ($f_{2.5, 2.8}=0$) or
under rotation. The values of $z$ and $\Qsat$ corresponding to the
circles (and pentagrams) can be read-off from Fig.~\ref{fig:MF_kepler}. 
The corresponding dimensionless spin parameters $\chi_{2.5, 2.8}$
($\chi \equiv J/M^2$ with $J$ being the angular momentum of the
pulsar) and strangeness fractions $F_{2.5, 2.8}$ for the same models
are shown in Figs.~\ref{fig:FFM_Rot}\,(b) and (c). Note that any
particular model is uniquely identified by their static maximum masses
$M^{\rm max}_{\rm TOV}$ shown by the horizontal axis. For the sake of
comparison, we show in Fig.~\ref{fig:FFM_Rot}\,(a) the frequency 716
Hz~\citep{Hessels:2006} of PSR J1748-2446ad, which has the highest
rotation frequency of all known pulsars.  In addition, we show the
rather speculative case of a possibly ultra-fast rotating object with
a frequency of 1250 Hz, suggested by the observation of narrow pulses
in the fast radio burst FRB 181112~\citep{Yamasaki:2020}.

The EoS models identified by their $M^{\rm max}_{\rm TOV}$ value
suggest the following comments on the possible origin of very massive
CS: (i) For $M^{\rm max}_{\rm TOV} \sim 2.1\,M_\odot$, the secondary
objects in the GW190814 event would need to be rotating at a frequency
$f_{2.5}\gtrsim 1200$~Hz, which is close to the Keplerian limit.  
(ii) For $M^{\rm max}_{\rm TOV} \sim 2.3\,M_\odot$, the secondary's
rotational frequency needs to be about 1000~Hz, which is below the
Keplerian limit and is by 25\% larger than that of PSR J1748-2446ad.
(iii) Finally, if  $M^{\rm max}_{\rm TOV} \sim 2.5\,M_\odot$, the
secondary of GW190814 is either a static or a slowly spinning CS, with
a frequency that is far below the Keplerian one and that of PSR J1748-2446ad.

Less can be said about the nature of the GW200210's secondary, due to
the large uncertainty in its mass. Nevertheless, the comments made
above about GW190814's secondary apply to the GW200210's secondary
too, provided its mass is $M_2 \simeq 2.5\,M_{\odot}$. Taking the larger mean
value $M_2 = 2.8\,M_{\odot}$ as a working hypothesis, one deduces that
$M^{\rm max}_{\rm TOV} \gtrsim 2.3\,M_\odot$ which would require
spin frequencies $f_{2.8}\gtrsim 1200$~Hz for the secondary to
be a CS. Qualitatively we may conclude that the above models require
stiff nucleonic EoS with $\Qsat \gtrsim 500$~MeV and maximally broken
SU(6) symmetry, see Figs.~\ref{fig:MF_static}\,(a) and (d).  As seen
from Fig.~\ref{fig:FFM_Rot}\,(b), the dimensionless spin parameters has values
$\chi_{2.5, 2.8} \lesssim 0.7$ for our models. The maximum value
$\chi_{\rm max} = 0.7$, which correspond to the Keplerian limit, is
essentially independent of the EoS models and is consistent with that
obtained in Refs.~\citep{Cook:1994,Haensel:1995,Lokw:2011}.  Finally, as
seen from Fig.~\ref{fig:FFM_Rot}\,(c), the CS models that can account
for very large masses contain a marginal of hyperons. 
For example, we find that the strangeness fraction is $F_{2.5}\lesssim 3\%$ for a
$M = 2.5\,M_\odot$ star and $F_{2.8}\lesssim 2\%$
for a $M = 2.8\,M_\odot$ star.
These values imply that massive stars are almost purely nucleonic.

\section{Summary and conclusions}
\label{sec:conclusion}
In this work, we constructed EoS models within CDF theory with degrees
of freedom that include the full baryon octet. The meson-hyperon
coupling constants are chosen to break the SU(6) spin-flavor symmetry
down to SU(3). The hyperon potentials were further fitted to the most
reliable values of their potentials at nuclear saturation density
extracted from hypernuclear data. Because of the more general SU(3) 
symmetry, the hyperonic couplings depend on additional parameters, among 
which the $z$-parameter (defined above) is most suitable for exploring the
impact of symmetry breaking. The density-dependences of the nucleonic and
hyperonic couplings is modeled using the same parameters.  The nucleonic
sector of the CDF was modeled phenomenologically at high density by
varying the slope coefficient $\Lsym$ and skewness coefficient $\Qsat$,
while maintaining the low-density features predicted by the DDME2
parametrization.

With this input, we investigated the mass and radius of non-rotating
as well as rapidly rotating stellar configurations. Our EoS models can
accommodate static CSs as massive as $M \simeq 2.3$-$2.5\,M_{\odot}$
in the large-$\Qsat$ and small-$z$ domain. However, the hyperon
content in this regime drops to several percent and, therefore, cannot
significantly influence the 
properties of CS. Thus, one may
conclude that the highly massive stellar models obtained with the SU(3) symmetric
models for the EoS are essentially nucleonic stars. 
The global parameters of these
stars are consistent with the parameters of stars based on 
purely nucleonic EoS 
models~\citep{Lijj:2020,Fattoyev:2020,Zhangnb:2020,Tsokaros:2020,Huangkx:2020,Biswas:2021,Tews:2021} (we exclude here the models calling for quark 
deconfinement~\citep{Dexheimer:2020,Tanhuang:2020,Bombaci:2020,Roupas:2020,Christian:2020,Demircik:2020,Malfatti:2020,Jumin:2021}).
This also confirms that genuinely hyperonic stars with a significant hyperonic fraction of 10-20\% are confined to lower
masses~\cite{Sedrakian:2020,Lijj:2020,Dexheimer:2020,Tuzh:2022,Liang:2022}.

We further constructed the rotating counterparts of our static stellar
models, including stars rotating at the Keplerian limit, in which case
the maximum mass of the nearly nucleonic models can reach values up to
$3.0\,M_{\odot}$. Our modeling allows us to estimate the ratio of the 
Keplerian to static maximum mass, $\eta$, and to show that it is constant 
only for stars with $M^{\rm max}_{\rm TOV} \ge 2.0\,M_{\odot}$. We find a
linear dependence of $\eta$ on $M^{\rm max}_{\rm TOV}$. Note that
sub-two-solar-mass stars are not excluded if there are two families of
CS in which the massive stars are strange.

We have also determined the minimum frequencies required to explain
the secondary stellar objects in the gravitational events GW190814 and
GW200210. We found that the most extreme models from the large-$\Qsat$
and small-$z$ domain produce masses in the required range.  This
domain is minimal for non-rotating stars and increases as rotation is
allowed.  It is maximal for the Keplerian case which allows for very
fast rotation at frequencies $f \sim 1500$~Hz.  We stress again that
even though our CDF study includes the full baryon octet, the highly
massive CS models turn out to contain only a very small amount of
hyperons (strangeness fractions of typical $\lesssim 3\%$). These
stars can therefore be considered as nucleonic stars.

\nolinenumbers
\section*{Acknowledgements}
The research of H.F. and J.L. is supported by the National Natural Science 
Foundation of China (Grant No. 12105232), the Venture \& Innovation Support 
Program for Chongqing Overseas Returnees (Grant No. CX2021007), and by the 
``Fundamental Research Funds for the Central Universities" (Grant No. SWU-020021).  
The research of A.S. is funded by Deutsche Forschungsgemeinschaft Grant 
No. SE 1836/5-2 and the Polish NCN Grant No. 2020/37/B/ST9/01937 at Wrocław
University.  F.W. acknowledges support by the U.S. National Science
Foundation under Grant PHY-2012152.


\end{document}